\documentclass[conference]{IEEEtran}
\IEEEoverridecommandlockouts
\usepackage{cite}
\usepackage{amsmath,amssymb,amsfonts}
\usepackage{algorithmic}
\usepackage{graphicx}
\usepackage{textcomp}
\usepackage{xcolor}
\usepackage{hyperref}
\usepackage{caption}
\usepackage{multirow}
\usepackage{subcaption}
\def\BibTeX{{\rm B\kern-.05em{\sc i\kern-.025em b}\kern-.08em
    T\kern-.1667em\lower.7ex\hbox{E}\kern-.125emX}}
\begin{document}

\title{Brain Tumor Diagnosis and Classification via Pre-Trained Convolutional Neural Networks}

\author{\IEEEauthorblockN{Dmytro Filatov}
\IEEEauthorblockA{\textit{Artificial Intelligence and Computer Vision} \\
\textit{Aimech Technologies Corp.}\\
San Francisco, The United States of America \\
dmytro@aimechtechnologies.com}
\and
\IEEEauthorblockN{Ghulam Nabi Ahmad Hassan Yar}
\IEEEauthorblockA{\textit{Department of Electrical and Computer Engineering} \\
\textit{Air University}\\
Islamabad, Pakistan \\
gnahy1@gmail.com}
}

\maketitle

\begin{abstract}
The brain tumor is the most aggressive kind of tumor and can cause low life expectancy if diagnosed at the later stages. Manual identification of brain tumors is tedious and prone to errors. Misdiagnosis can lead to false treatment and thus reduce the chances of survival for the patient. Medical resonance imaging (MRI) is the conventional method used for the diagnosis of brain tumors and their types. This paper attempts to eliminate the manual process from the process of diagnosis and use machine learning instead. We proposed the use of pre-trained convolutional neural networks (CNN) for the diagnosis and classification of brain tumors. Three types of tumors were classified with one class of non-tumor MRI images. Networks that has been used are ResNet50, EfficientNetB1, EfficientNetB7, EfficientNetV2B1. EfficientNet has shown promising results due to its scalable nature. EfficientNetB1 showed the best results with training and validation accuracy of 87.67\% and 89.55\% respectively.
\end{abstract}


\section{Introduction}

The brain is the most important organ of the body which controls the other organs. A tumor is the abnormal growth of the cells due to the uncontrolled division of cells. Tumors are divided into two grades, low and high. Low-grade tumors are benign and are not cancerous, while high-grade tumors are malignant and can spread to other parts of the body and can cause death \cite{seetha2018brain}. In 2016 brain tumor was the leading cause of death in children. Brain tumors are also the third most commonly occurring cancer among people between ages 15 and 39 \cite{abiwinanda2019brain}. Different type of tumors needs different type of treatments and misdiagnosis can cause the death of the patient. Tumors are diagnosed on basis of their size, location, and intensity. This is good if the surgery is needed but for treatment, the type of tumor plays a crucial role.

Medical Resonance Imaging (MRI), Computed Tomography (CT), and Ultrasound can be used to diagnose the tumor. MRI has shown more promising results than the other two kinds of radiology methods. In MRI the brain tumor can be spotted as the most bright part. MRI generated an image on basis of the number of hydrogen atoms in the body. If the number of hydrogen atoms is high in an area, then the area appears to be bright. In the brain, cerebrospinal fluid (CSF) is the area that has more number of hydrogen atoms and it appears brighter than the other areas. Other than CRF, tumors have the highest number of hydrogen atoms so it also appears as bright areas. If the tumor is not visible clearly with a normal MRI then a contrast agent can help to highlight the tumor. Gadolinium is the contrast agent that is used in MRI. Gadolinium has a high quantity of glucose and glucose has a high quantity of hydrogen. Tumors are known to absorb the glucose, when tumors do so they become rich in hydrogen and thus appear brighter in the MRI images.

\begin{figure}
    \centering
    \includegraphics[width=\linewidth]{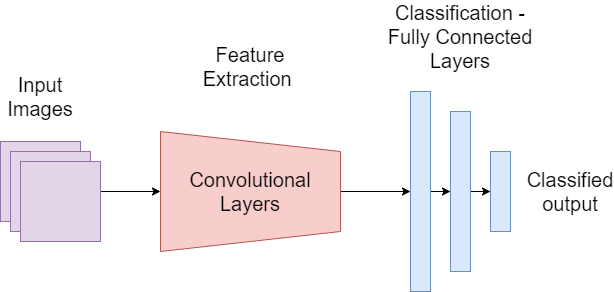}
    \caption{Division of CNN architecture}
    \label{fig:cnn_div}
\end{figure}

After the MRI of the brain is taken features can be extracted for automated classification of the tumor. These features can be extracted via Bag-of-Words (BoW) or Fisher Vector. After the feature extraction, classification models can be used to classify those features thus classifying the type of tumor. Convolutional neural networks are the combination of feature extractors and classifiers. Features are extracted through the convolutional layers and classification happens using the fully connected layers on basis of extracted features. The basic architectural diagram of CNN can be seen in \autoref{fig:cnn_div}.

Transfer learning is the concept where the weights of an already trained model on some dataset are used for some other application. For example, for the problem under consideration in this research, weights were used from the model that was trained on the ImageNet dataset for 1000 classes. This use of pre-trained models helps in faster convergence and the model can be trained on fewer resources and fewer datasets.

This paper proposes the correct diagnosis and classification of brain tumors through MRI images by using pre-trained CNN. Pre-trained models that has been used for comparison are EfficientNetB1, EfficientNetB7, EfficientNetV2B1, and ResNet50. This classification can help incorrect treatment of the tumors. The goals achieved by this paper are:

\begin{enumerate}
    \item Diagnosis of brain tumor.
    \item Classification of three kinds of tumors: Glioma, Meningioma, and Pituitary.
    \item Use of pre-trained models to reduce the resources used and mitigate the effect of small dataset.
\end{enumerate}

The rest of the paper is divided as follows: Section 2 will present the work that has already been done in the diagnosis, segmentation, and classification of brain tumors. This section will discuss the methods used, and the results that have been achieved by using those methods. Section 3 will discuss the methodology used for the diagnosis and classification of the tumors. This section will put light on pre-processing that has been applied to the input dataset for better classification, modifications made to the architecture of fully connected layers, and how implementation was done. Section 4 will discuss the dataset and its division into training and validation datasets. Section 5 will be about the presentation and discussion of the results. This section will also conclude the paper and proposed some future changes that can improve the quality of the diagnosis and classification. 

\section{Literature Review}

A lot of literature is already present but most of the work uses the old models. Deepak and Ameer \cite{deepak2019brain} worked on classifying three kinds of brain tumors that are glioma, meningioma, and pituitary. They used GoogleNet using transfer learning for the classification. Seetha and Raja \cite{seetha2018brain} highlighted the severeness of brain tumors and the importance of correct classification of brain tumors. They used CNN for the tumor classification. Abiwinanda et al. \cite{abiwinanda2019brain} stated that manually diagnosing the class of tumor is a tiresome process and has a margin of error. They worked on the classification of glioma, meningioma, and Pituitary tumors using pre-trained models. They tested five different ANN architectures in front of convolutional layers. Sajjad et al. \cite{sajjad2019multi} worked on the multi-grade classification of brain tumors. They first segmented the brain tumor part of the brain using deep learning and then used fine-tuned pre-trained CNN models for classification. Khan et al. \cite{khan2020multimodal} worked on multi-modal brain tumor classification. By multi-modal they mean that Medical Resonance Images (MRI) are taken via different methods that are T1, T2, T1CE, and Flair. They used VGG16 and VGG19 CNN for feature extraction and in the end, using an extreme learning machine (ELM) for final classification. Swati et al. \cite{swati2019brain} highlighted that in brain tumor classification either low-level features or high-level features are used. Some researchers used hand-crafted features for better classification. Recent deep learning models have shown good classification, but they require large datasets. To work with small datasets, they proposed the use of pre-trained models. MRI images used by then were T1-weighted contrast-enhanced. Swati et al. \cite{swati2019content} in another research used VGG-19 based feature extractor along with closed-form metric learning for finding similarities between query images and database images. Sharif et al. \cite{sharif2021decision} highlighted that brain tumor classification is a crucial task and yet present research suffers from low accuracy. To solve this problem, they proposed a combination of multiple feature extraction methods. The first method they used was Entropy–Kurtosis-based High Feature Values (EKbHFV) and the second was a modified genetic algorithm (MGA) based on metaheuristics. In the end, classification was done by an SVM cubic classifier. Ari and Hanbay \cite{ari2018deep} proposed a three-step pipeline for brain tumor classification and detection. The first step was pre-processing, in which local smoothing and nonlocal smoothing methods were used to remove noise. The second step was to use the extreme learning machine local receptive fields (ELM-LRF) for brain tumor classification.  In the last step, they used image processing for brain tumor extraction. Afshar et al. \cite{afshar2019capsule} highlighted an important feature for the classification of brain tumors. They stated that the relation between a tumor and its surrounding tissues is of great importance for brain tumor classification. CNN failed to fully utilize the spatial features that are important for this classification. They proposed the use of CapsNets for brain tumor classification. The drawback they highlighted in the use of CapsNets was that it is sensitive to the background. So, additional input of the area of the tumor was given to the CapsNets for better classification. Chelghoum et al. \cite{chelghoum2020transfer} collected a benchmark dataset of contrast-enhanced MRI images. They worked on the classification of three kinds of brain tumors  (glioma, meningioma, and pituitary). As they had a small dataset they used pre-trained models for classification. They trained the model for the different number of epochs to see the relationship between time and model accuracy. Their results showed that the model showed acceptable results even with a few epochs and a small dataset when pre-trained models are used. Mehrotra et al. \cite{mehrotra2020transfer} worked on the binary classification of tumor malignant or benign. They used pre-trained models as the dataset was small and has only 696 T1-weighted MRI images. Khan et al. \cite{khan2020brain} used VGG-16, ResNet-50, and Inception-v3 models for brain tumor classification using MRI images.

There has been a lot of literature present on brain tumor classification and there is more to come as the problem has not been fully solved with non-challengeable accuracy.

\section{Methodology}
\begin{figure*}
    \centering
    \includegraphics[width=\linewidth]{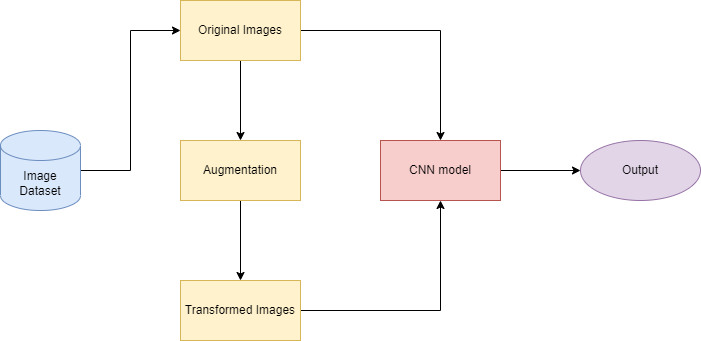}
    \caption{Classification pipeline}
    \label{fig:pipe}
\end{figure*}

The overall pipeline that has been used is to first apply image augmentation on the input images to increase the size of input data. Then the pre-trained convolutional layers have been used to extract the features from those images. Then fully connected artificial neural network designed by us has been used to classify the extracted features thus classifying the input MRI images of the brain. \autoref{fig:pipe} shows the overall classification pipeline.

\subsection{Pre-processing}

\begin{figure}
    \centering
    \includegraphics[width=\linewidth]{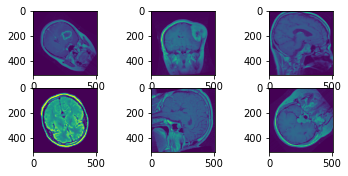}
    \caption{A few examples of image augmentation}
    \label{fig:aug}
\end{figure}

The pre-processing step includes the augmentation of the images to mitigate the effect of fewer input images. Augmentation is the process of taking an image and generating its variants by translations, rotations, scaling, shearing, and flipping (horizontal and vertical). The machine learning model treats these variants as a different image and thus the size of the dataset increases. \autoref{fig:aug} shows a few variants of the images generated by image augmentation. Augmentation used in this research includes rotation of $90^{0}$, horizontal flip, and vertical flip.

\subsection{Model Architecture}

\begin{figure}
    \centering
    \includegraphics[width = \linewidth]{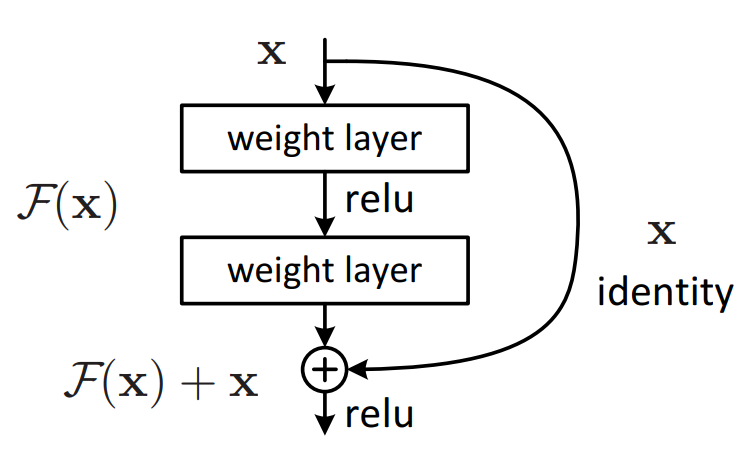}
    \caption{Residual block, basic architectural block of ResNet \cite{he2016deep}}
    \label{fig:resnet}
\end{figure}

After pre-processing and image augmentation, classification models were implemented. Classification models used were ResNet50, and EfficientNet. ResNet50 was the model proposed by Microsoft in 2015 \cite{he2016deep}. The backbone of this model is the residual block that used the identity mapping. \autoref{fig:resnet} shows the residual block of ResNet model. Identity mapping is to use the input from previous layers without any change, This helps in tackling the problem of vanishing gradient. Another advantage of the residual block is that much deeper networks can be built without increasing the percentage of training error.

\begin{figure*}
    \centering
    \includegraphics[width = \linewidth]{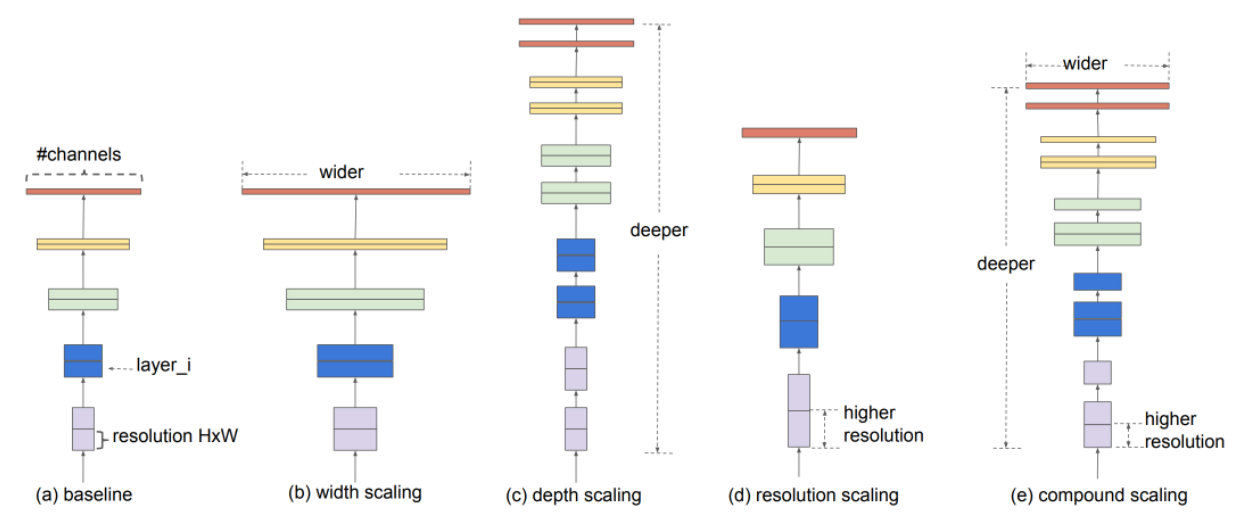}
    \caption{EfficientNet: (a) Basic architecture, (b) Width scaling, (c) Depth scaling, (d) Resolution scaling (e) Compound scaling \cite{tan2019efficientnet}}
    \label{fig:efficient}
\end{figure*}

EffieientNet \cite{tan2019efficientnet} on the other hand does not go into depth but also introduces the width scaling, and resolution scaling. EfficientNet scales all the three dimensions at a fixed ratio. EfficientNets were introduced by the Google research team in 2019. They showed the performance improvement by first introducing a base CNN and then applying one scaling method at a time. In the end, all the scaling methods were implemented and a performance improvement was observed. \autoref{fig:efficient} shows different kind of scalings that has been used in EfficientNet along with compound scaling which is EfficientNet.

For the classification, fully connected layers were to be used. As the pre-trained model architectures used has 1000 output classes, it can not be used for the problem under consideration as there are only four classes to be classified. So, a new architecture was defined consisting of pooling, flattening, dense, and dropout layers. The architecture used is shown in \autoref{tab:ann} \cite{albahli2021fast}.

\begin{table}[]
    \centering
    \caption{Architecture of fully connected layers \cite{albahli2021fast}}
    \begin{tabular}{lll}
\hline
Layer           & Output                 & kernel     \\ \hline
Average Pooling & 4$\times$4$\times$2048 & 2$\times$2 \\
Flatten         & 32768                  & N/A$^{a}$  \\
Dense           & 1024                   & N/A        \\
Dropout (0.5)   & 1024                   & N/A        \\
Dense           & 1024                   & N/A        \\
Dropout (0.5)   & 1024                   & N/A        \\
Dense           & 4                      & N/A        \\ \hline
\multicolumn{3}{l}{$^{a}$N/A: not applicable}         \\ \hline
\end{tabular}
\label{tab:ann}
\end{table}

\subsection{Implementation}

Models were trained on Google Colab with 12.5GB of RAM and Tesla P100-PCIE-16GB GPU. The size of input images was kept 512$\times$512$\times$3. During the training accuracy and loss were monitored. The learning rate was kept at 0.001 with a batch size of 8. The loss function used was categorical cross-entropy. To minimize this function Adam optimizer was used. Data were split into training and validation at a ratio of 80\% and 20\%. During the training validation loss was monitored to save the model and apply early stopping if necessary. Model weights were stored every time the minimum value of validation loss was achieved. The maximum number of epochs was set to 50 and if the validation loss did not decrease for 9 consecutive epochs then the model training was stopped.

\section{Dataset}

Dataset used in this research was taken from Kaggle database\footnote{https://www.kaggle.com/datasets/masoudnickparvar/brain-tumor-mri-dataset}. The dataset is a combination of data from three other databases figshare\footnote{https://figshare.com/articles/dataset/brain\_tumor\_dataset/1512427}, SARTAJ dataset\footnote{https://www.kaggle.com/datasets/sartajbhuvaji/brain-tumor-classification-mri}, Br35H\footnote{https://www.kaggle.com/datasets/ahmedhamada0/brain-tumor-detection}. Dataset is taken from different sources so they had different resolutions. The SARTAJ dataset has mislabeled entries in the glioma class, so images from the glioma class were not added to the database. In total there are 7022 images of human brain MRI. These images were divided into training and testing datasets at a ratio of 80\% and 20\%. \autoref{tab:data_div} shows the division in terms of numbers.

\begin{table}[]
    \centering
    \caption{Dataset distribution into training and testing images}
    \begin{tabular}{|l|l|l|l|l|}
\hline
Type       & Glioma & Meningioma & Pituitary & Normal \\ \hline
Total      & 1621   & 1645       & 1757      & 2000   \\ \hline
Training   & 1321   & 1339       & 1457      & 1595   \\ \hline
Validation & 300    & 306        & 300       & 405    \\ \hline
\end{tabular}
    
    \label{tab:data_div}
\end{table}

\begin{figure*}
    \centering
    \includegraphics[width=\columnwidth]{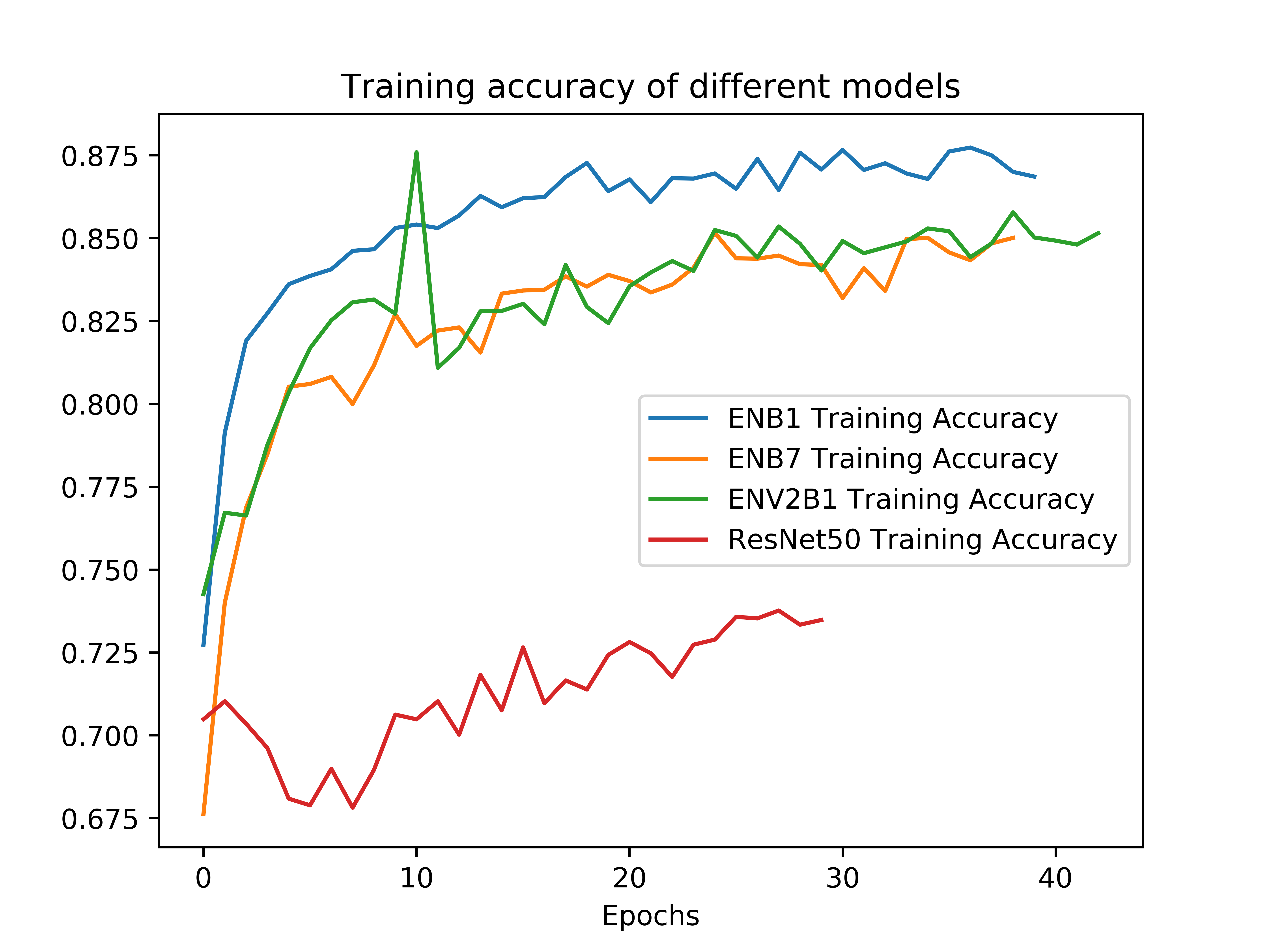}
    \includegraphics[width=\columnwidth]{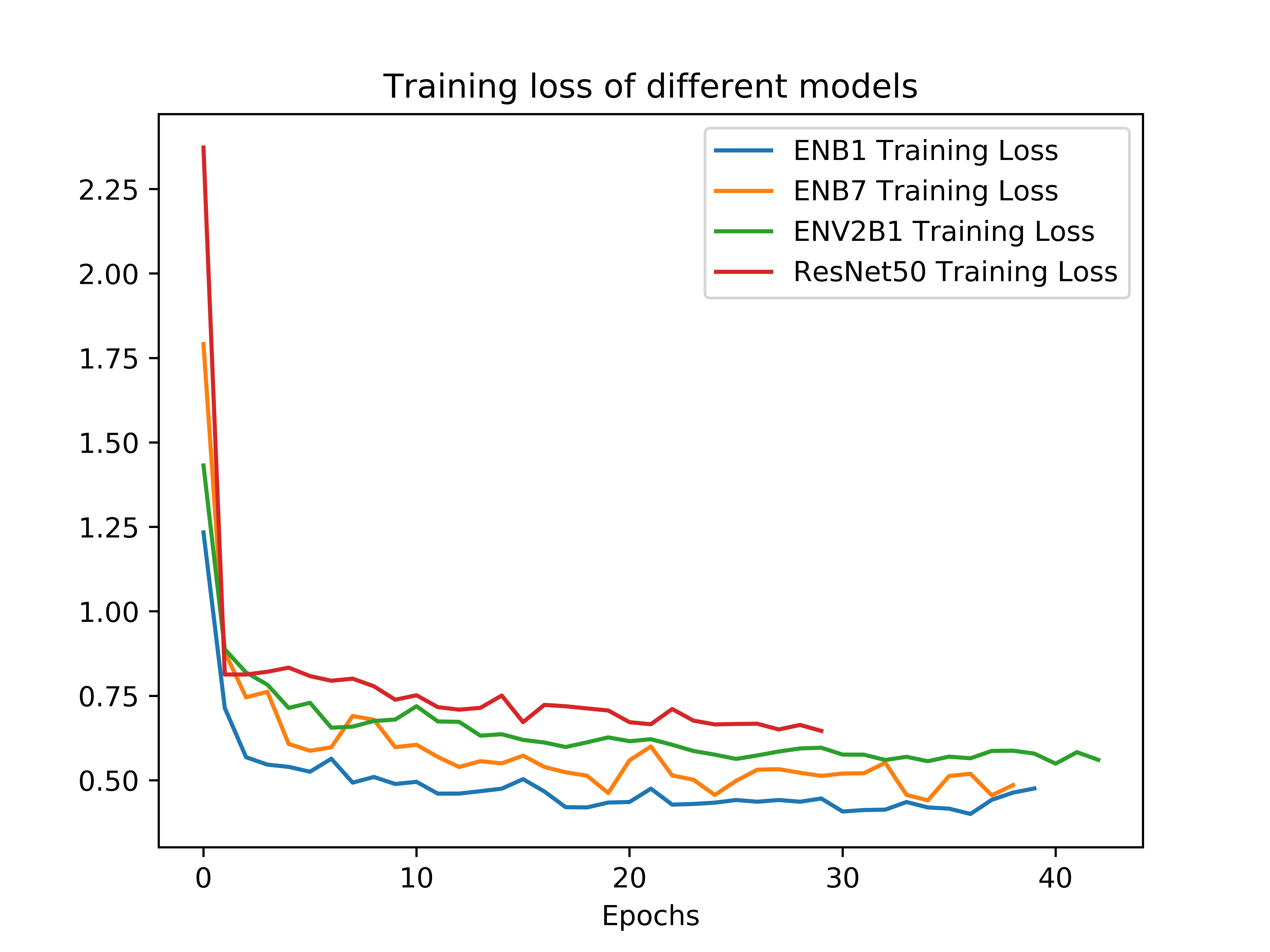}
    \caption{Training accuracy (left) and loss (right) for different models}
    \label{fig:train}
\end{figure*}

\begin{figure*}
    \centering
    \includegraphics[width=\columnwidth]{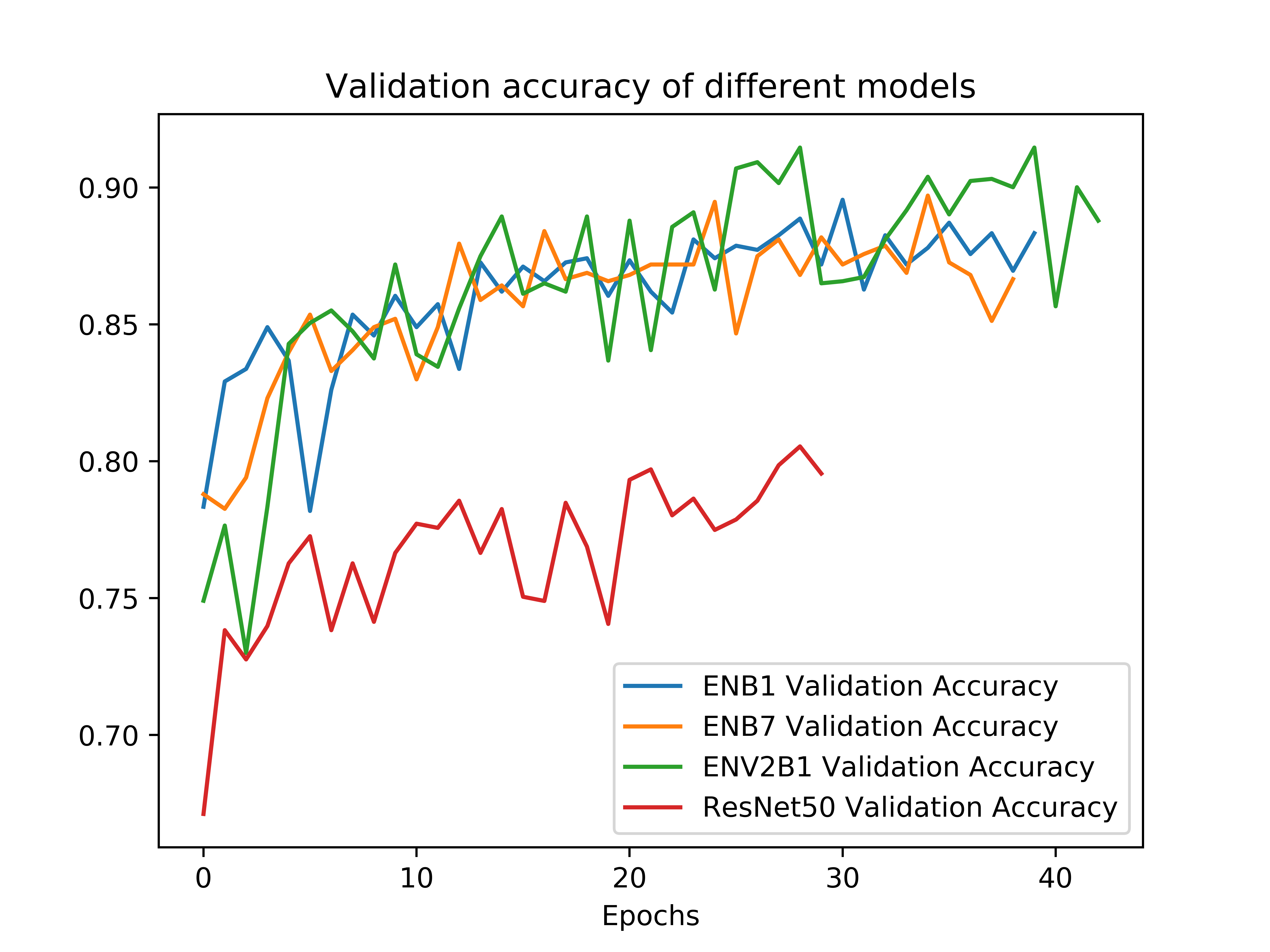}
    \includegraphics[width=\columnwidth]{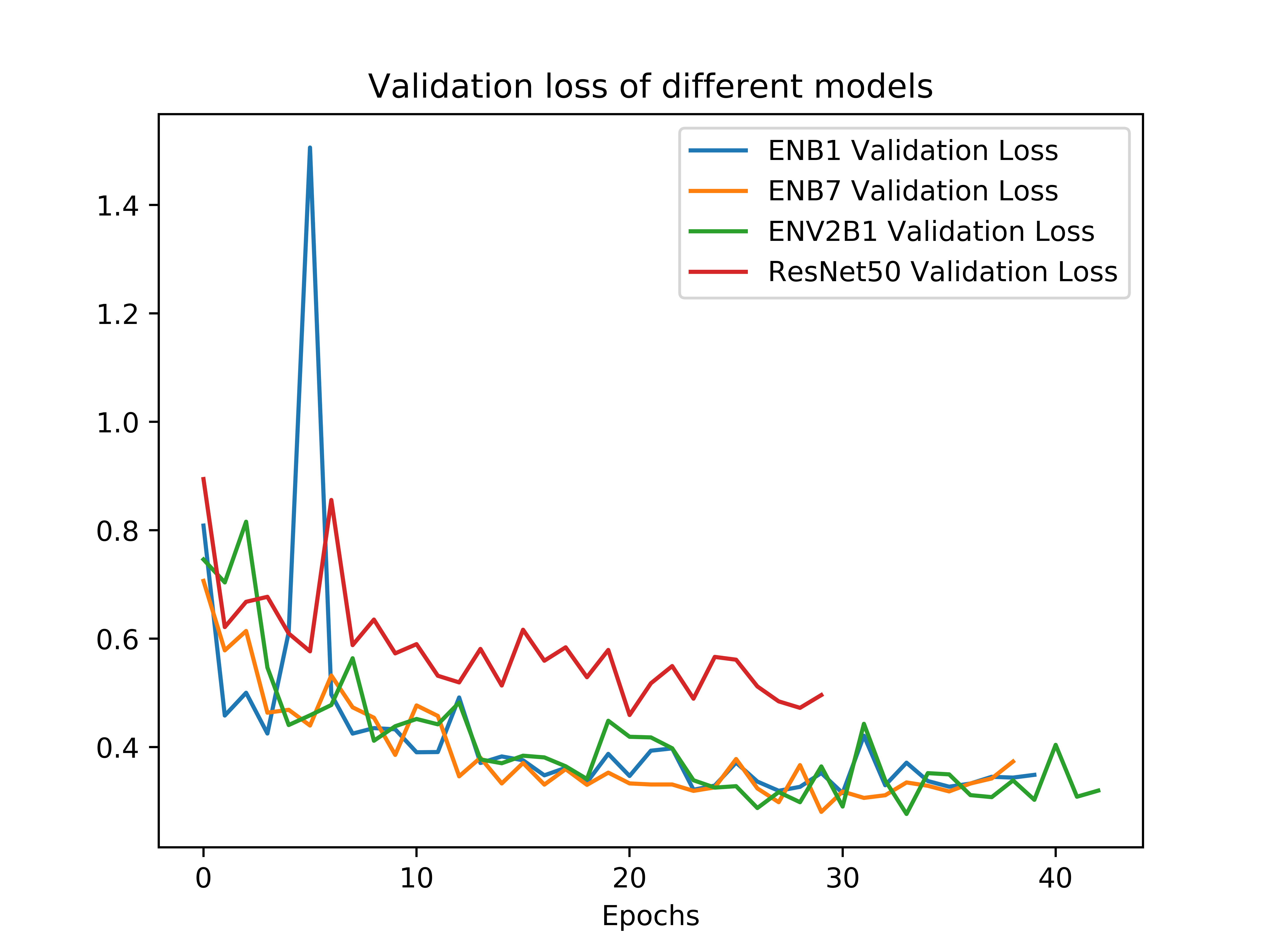}
    \caption{Validation accuracy (left) and loss (right) for different models}
    \label{fig:val}
\end{figure*}

\section{Results \& Discussion}

As explained above the first step in the machine learning pipeline for brain tumor classification is the reshaping and augmentation of the dataset. \autoref{fig:aug} shows the augmentation results to increase the size of the input dataset. After the image augmentation classification was implemented using different pre-trained models. \autoref{tab:res} shows the classification results of different datasets in terms of training and validation accuracy and loss. From the table, it is clear that the EfficientNet models surpassed the ResNet models with a significant margin. It can also be noted that the validation loss stopped decreasing after epoch 21 because the total number of the epoch was 30 for ResNet50. While for other models validation loss stopped decreasing at later stages. This also explains the high validation loss of the ResNet50 model. EfficientNetB1 showed the best performance with the best training and validation accuracy. Training loss was also minimum for EfficientNetB1 but validation loss was the third lowest.

\autoref{fig:train} shows the training accuracy and loss for all the models. All EfficientNet models showed a steady increase in training accuracy and a steady decrease in training loss. While training accuracy for the ResNet50 model was not stable as it decreased at first and then increased but the difference between initial and final accuracy was not large. 

The same results can be seen in terms of validation accuracy and loss in \autoref{fig:val}. All models showed promising results but ResNet50 remained behind them. Although a spike in validation loss can be seen for the EfficientNetB1 model at epoch 6, the model showed lower validation loss after that epoch. This behavior is not new when considering the validation loss. The reason is that validation data is unseen to the model and sometimes a batch of data appears that is difficult to predict.

\begin{table}[]
    \centering
    \caption{Classification results for different models}
    \begin{tabular}{|l|l|l|l|l|l|}
\hline
\textit{\textbf{Model}}            & \textit{\textbf{Epochs}} & \textit{\textbf{\begin{tabular}[c]{@{}l@{}}Training \\ Accuracy\end{tabular}}} & \textit{\textbf{\begin{tabular}[c]{@{}l@{}}Training \\ Loss\end{tabular}}} & \textit{\textbf{\begin{tabular}[c]{@{}l@{}}Val \\ Accuracy\end{tabular}}} & \textit{\textbf{\begin{tabular}[c]{@{}l@{}}Val\\  Loss\end{tabular}}} \\ \hline
\textit{\textbf{EfficientNetB7}}   & 39                      & 0.8419                                                                         & 0.5129                                                                     & 0.8818                                                                    & 0.2807                                                                \\ \hline
\textit{\textbf{EfficientNetV2B1}} & 43                      & 0.8491                                                                         & 0.5695                                                                     & 0.8917                                                                    & 0.2768                                                                \\ \hline
\textit{\textbf{EfficientNetB1}}   & 40                      & 0.8767                                                                         & 0.4076                                                                     & 0.8955                                                                    & 0.3152                                                                \\ \hline
\textit{\textbf{ResNet50}}         & 30                      & 0.7282                                                                         & 0.6719                                                                     & 0.7932                                                                    & 0.4593                                                                \\ \hline
\end{tabular}
    \label{tab:res}
\end{table}

\subsection{Conclusion \& Future Recommendations}

This paper presents an application of machine learning in the domain of medical imaging for brain tumor detection and classification. Image augmentation and transfer learning was used to mitigate the effects of small dataset and computational power. The performance of different models was discussed and it was shown that EfficientNet models are more effective than any other model due to their width, depth, and resolution scaling properties. EfficientNetB1 showed the best performance in terms of both training and validation accuracy.

For future work, it is suggested to collect more data for better classification accuracy. Other models can also be trained for comparison i.e., EfficientNetV2L, EfficientNetV2B2, and EfficientNetV2B3, etc., and the best model can be selected. Another change can be done in the ANN part of the neural network. Only one architecture was tested in this paper and multiple architectures can be designed and checked for better accuracy.

\bibliographystyle{IEEEtran}
\bibliography{ref}

\begin{thebibliography}{10}
\providecommand{\url}[1]{#1}
\csname url@samestyle\endcsname
\providecommand{\newblock}{\relax}
\providecommand{\bibinfo}[2]{#2}
\providecommand{\BIBentrySTDinterwordspacing}{\spaceskip=0pt\relax}
\providecommand{\BIBentryALTinterwordstretchfactor}{4}
\providecommand{\BIBentryALTinterwordspacing}{\spaceskip=\fontdimen2\font plus
\BIBentryALTinterwordstretchfactor\fontdimen3\font minus
  \fontdimen4\font\relax}
\providecommand{\BIBforeignlanguage}[2]{{%
\expandafter\ifx\csname l@#1\endcsname\relax
\typeout{** WARNING: IEEEtran.bst: No hyphenation pattern has been}%
\typeout{** loaded for the language `#1'. Using the pattern for}%
\typeout{** the default language instead.}%
\else
\language=\csname l@#1\endcsname
\fi
#2}}
\providecommand{\BIBdecl}{\relax}
\BIBdecl

\bibitem{seetha2018brain}
J.~Seetha and S.~S. Raja, ``Brain tumor classification using convolutional
  neural networks,'' \emph{Biomedical \& Pharmacology Journal}, vol.~11, no.~3,
  p. 1457, 2018.

\bibitem{abiwinanda2019brain}
N.~Abiwinanda, M.~Hanif, S.~T. Hesaputra, A.~Handayani, and T.~R. Mengko,
  ``Brain tumor classification using convolutional neural network,'' in
  \emph{World congress on medical physics and biomedical engineering
  2018}.\hskip 1em plus 0.5em minus 0.4em\relax Springer, 2019, pp. 183--189.

\bibitem{deepak2019brain}
S.~Deepak and P.~Ameer, ``Brain tumor classification using deep cnn features
  via transfer learning,'' \emph{Computers in biology and medicine}, vol. 111,
  p. 103345, 2019.

\bibitem{sajjad2019multi}
M.~Sajjad, S.~Khan, K.~Muhammad, W.~Wu, A.~Ullah, and S.~W. Baik, ``Multi-grade
  brain tumor classification using deep cnn with extensive data augmentation,''
  \emph{Journal of computational science}, vol.~30, pp. 174--182, 2019.

\bibitem{khan2020multimodal}
M.~A. Khan, I.~Ashraf, M.~Alhaisoni, R.~Dama{\v{s}}evi{\v{c}}ius, R.~Scherer,
  A.~Rehman, and S.~A.~C. Bukhari, ``Multimodal brain tumor classification
  using deep learning and robust feature selection: A machine learning
  application for radiologists,'' \emph{Diagnostics}, vol.~10, no.~8, p. 565,
  2020.

\bibitem{swati2019brain}
Z.~N.~K. Swati, Q.~Zhao, M.~Kabir, F.~Ali, Z.~Ali, S.~Ahmed, and J.~Lu, ``Brain
  tumor classification for mr images using transfer learning and fine-tuning,''
  \emph{Computerized Medical Imaging and Graphics}, vol.~75, pp. 34--46, 2019.

\bibitem{swati2019content}
------, ``Content-based brain tumor retrieval for mr images using transfer
  learning,'' \emph{IEEE Access}, vol.~7, pp. 17\,809--17\,822, 2019.

\bibitem{sharif2021decision}
M.~I. Sharif, M.~A. Khan, M.~Alhussein, K.~Aurangzeb, and M.~Raza, ``A decision
  support system for multimodal brain tumor classification using deep
  learning,'' \emph{Complex \& Intelligent Systems}, pp. 1--14, 2021.

\bibitem{ari2018deep}
A.~Ari and D.~Hanbay, ``Deep learning based brain tumor classification and
  detection system,'' \emph{Turkish Journal of Electrical Engineering \&
  Computer Sciences}, vol.~26, no.~5, pp. 2275--2286, 2018.

\bibitem{afshar2019capsule}
P.~Afshar, K.~N. Plataniotis, and A.~Mohammadi, ``Capsule networks for brain
  tumor classification based on mri images and coarse tumor boundaries,'' in
  \emph{ICASSP 2019-2019 IEEE International Conference on Acoustics, Speech and
  Signal Processing (ICASSP)}.\hskip 1em plus 0.5em minus 0.4em\relax IEEE,
  2019, pp. 1368--1372.

\bibitem{chelghoum2020transfer}
R.~Chelghoum, A.~Ikhlef, A.~Hameurlaine, and S.~Jacquir, ``Transfer learning
  using convolutional neural network architectures for brain tumor
  classification from mri images,'' in \emph{IFIP International Conference on
  Artificial Intelligence Applications and Innovations}.\hskip 1em plus 0.5em
  minus 0.4em\relax Springer, 2020, pp. 189--200.

\bibitem{mehrotra2020transfer}
R.~Mehrotra, M.~Ansari, R.~Agrawal, and R.~Anand, ``A transfer learning
  approach for ai-based classification of brain tumors,'' \emph{Machine
  Learning with Applications}, vol.~2, p. 100003, 2020.

\bibitem{khan2020brain}
H.~A. Khan, W.~Jue, M.~Mushtaq, and M.~U. Mushtaq, ``Brain tumor classification
  in mri image using convolutional neural network,'' \emph{Math. Biosci. Eng},
  vol.~17, no.~5, pp. 6203--6216, 2020.

\bibitem{he2016deep}
K.~He, X.~Zhang, S.~Ren, and J.~Sun, ``Deep residual learning for image
  recognition,'' in \emph{Proceedings of the IEEE conference on computer vision
  and pattern recognition}, 2016, pp. 770--778.

\bibitem{tan2019efficientnet}
M.~Tan and Q.~Le, ``Efficientnet: Rethinking model scaling for convolutional
  neural networks,'' in \emph{International conference on machine
  learning}.\hskip 1em plus 0.5em minus 0.4em\relax PMLR, 2019, pp. 6105--6114.

\bibitem{albahli2021fast}
S.~Albahli and G.~N. A.~H. Yar, ``Fast and accurate detection of covid-19 along
  with 14 other chest pathologies using a multi-level classification: Algorithm
  development and validation study,'' \emph{Journal of Medical Internet
  Research}, vol.~23, no.~2, p. e23693, 2021.

\end{thebibliography}

\end{document}